\documentclass[prletter,showpacs,twocolumn,typeset,superscriptaddress]{revtex4}
\usepackage{amsfonts}
\usepackage{amsmath}
\usepackage{amssymb}
\usepackage{graphicx}
\usepackage{epsfig}
\usepackage{dcolumn}
\usepackage{bm}

\begin{document}
\preprint{APS}

\title{Conclusive inner product modification}
\author{Luis Roa}
\affiliation{Center for Quantum Optics and Quantum Information,
Departamento de F\'{\i}sica, Universidad de Concepci\'{o}n, Casilla
160-C, Concepci\'{o}n, Chile.}

\author{A. Delgado}
\affiliation{Center for Quantum Optics and Quantum Information,
Departamento de F\'{\i}sica, Universidad de Concepci\'{o}n, Casilla
160-C, Concepci\'{o}n, Chile.}

\author{M. L. Ladr\'on de Guevara}
\affiliation{Departamento de F\'{\i}sica, Universidad Cat\'{o}lica
del Norte, Casilla 1280, Antofagasta, Chile.}

\date{\today}

\begin{abstract}
The task of changing the overlap between two quantum states
can not be performed by making use of a unitary evolution only.
However, by means of a unitary-reduction process it can be
probabilistically modified. Here we study in detail the problem of
mapping two known pure states onto other two states in such
a way that the final inner product between the outcome states is
different from the inner product of the initial states. In this way we design an optimal \textit{non-orthogonal
quantum state preparation} scheme by starting from an orthonormal
basis. In this scheme the absolute value of the inner product
can be reduced only probabilistically whereas it can be increased
deterministically. Our analysis shows that the phases of the involved inner products play an important role in the
increase of the success probability of the desired process.
\end{abstract}

 \maketitle

\section{Introduction}

The discrimination or identification of quantum states has been a
constant subject of study. In recent years, the interest for this
matter has been renewed due to the potential possibility of encoding
and transmitting information using quantum systems \cite{1.1}. In
addition, the current progress in experimental techniques has led to
physical implementations of quantum information protocols aimed at
identifying or discriminating quantum states \cite{1.2,1.3,1.4}.

A particular strategy for discriminating among quantum states is the
so called \emph{unambiguous state discrimination}
\cite{2,2.1,2.2,2.3,2.4,2.5,2.6}. This permits identifying
conclusively states belonging to a set of linearly independent
states \cite{2.4} with a success probability different from one.
This strategy is based on the concatenation of a unitary
transformation and two projective measurements. The unitary
transformation acts onto both the primary system, which encodes the
states to be discriminated, and an ancillary system. This
transformation is designed in such a way that a measurement on the
ancillary system projects with a given probability the primary
system onto one of several possible mutually orthogonal states.
Thereby, a measurement of the primary system leads to the perfect
identification of the state \cite{2.2,4}. Unambiguous state
discrimination has been proposed as basis for quantum key sharing
\cite{4.1} and as a tool for improving the quality of several
quantum protocols when implemented via partially entangled states
such as quantum teleportation \cite{5}, dense coding \cite{6}, and
entanglement swapping \cite{7}, concentration \cite{7.1}, and
generation \cite{7.2}.

Unambiguous state discrimination can also be viewed as a
probabilistic conclusive mapping connecting a set of initial states
with non-vanishing inner products with a set of final states with
vanishing inner products. One can think in a more general process
consisting in a probabilistic conclusive mapping which connects two
sets corresponding to initial and final states, each of them with
different inner products. This general mapping is the subject of
this article. Recently, the mapping between sets of non-orthogonal
states has been connected to the control of quantum state
preparation, entanglement modification \cite{9}, and interference in
the quantum eraser \cite{9.1}. In our study, we restrict ourselves
to the case of sets of two states and we allow different a priori
probabilities of the two initial states. The complex nature of the
inner product between them is also taken into account. We find
conditions that guarantee the existence of the mapping and we
optimize its success probability. The mathematical treatment of the
problem suggests to consider three different cases, which have to be
analyzed separately. These cases are classified according to
conditions satisfied by the absolute values and phases of the
involved inner products. One of the analyzed situations corresponds
to a scheme of optimal \textit{non-orthogonal quantum state
preparation} by starting from an orthonormal basis. We explore some
applications of the mapping to Quantum Information Theory, such as
quantum deleting and quantum cloning. We compare the optimal success
probability of our scheme with the one obtained  via an alternative
scheme using the unambiguous quantum state discrimination.

This article is organized as follows: in section \ref{section2} we introduce the mapping and we classify three possible solutions, which are analyzed in detail in three subsections. In section \ref{sectionA} we compare this with the alternative scheme.
In section \ref{application} we discuss some possible applications of the mapping, and finally in section \ref{summary} we summarize our results.

\section{mapping between sets of non-orthogonal states}
\label{section2}

Let us consider a primary quantum system $s$ described by a two-dimensional Hilbert space ${\cal H}_s$. This system is prepared randomly in one of the states $|\alpha_i\rangle_s$ ($i=1,2$),
with probability $\eta_i$, and then
it is delivered to a third party. This third party has a complete knowledge of the states $\{|\alpha_i\rangle_s\}$ and its a priory probabilities, but it does not known the actual state in which the system is. The task of this party is to map the states $\{|\alpha_i\rangle_s\}$ onto the final states $\{|\beta_i\rangle_s\}$, where the initial inner product $\langle \alpha_1|\alpha_2\rangle=\alpha$ is in general different from the inner product $\langle \beta_1|\beta_2\rangle=\beta$. The mapping must be implemented conclusively, that is, it must be known with certainty when it has been perfectly carried out, and without revealing the initial state of the primary system $s$. Furthermore, the mapping must achieve the highest possible success probability.

In order to implement the mapping we consider an ancillary quantum system $a$ described by a Hilbert space ${\cal H}_a$ spanned by the orthogonal states $\{|0\rangle_a,|1\rangle_a\}$. This allows by means of a joint unitary transformation and a measurement process to project indirectly states $\{|\alpha_i\rangle_s\}$ onto states $\{|\beta_i\rangle_s\}$ in a probabilistic way. We assume that there exists the joint unitary operator $U_{sa}$ such that
\begin{subequations}
\begin{eqnarray}
U_{sa}|\alpha_1\rangle_s|A\rangle_a=\sqrt{p_1}|\beta_1\rangle_s|0\rangle_a+\sqrt{1-p_1}|\gamma_1\rangle_s|1\rangle_a, \\
U_{sa}|\alpha_2\rangle_s|A\rangle_a=\sqrt{p_2}|\beta_2\rangle_s|0\rangle_a+\sqrt{1-p_2}|\gamma_2\rangle_s|1\rangle_a,
\end{eqnarray} \label{TRANSFORMATION}%
\end{subequations}%
where the ancillary system $a$ is initially prepared in a known, arbitrary state $|A\rangle_a$. Since the states $\{|0\rangle_a,|1\rangle_a\}$ are orthonormal, a measurement of the ancilla on this basis projects the initial state $|\alpha_i\rangle_s$ onto the state $|\beta_i\rangle_s$ with probability $p_i$ or onto the state $|\gamma_i\rangle_s$ with probability $1-p_i$ ($i=1,2$). Thus, the total success probability $P_\beta$ of mapping the set of states $\{|\alpha_i\rangle_s\}$ onto the set of states $\{|\beta_i\rangle_s\}$ is given by
\begin{equation}
P_\beta=\eta_1p_1+\eta_2p_2.
\label{pbeta}
\end{equation}
Analogously, the total success probability $P_\gamma$ of mapping the set of states $\{|\alpha_i\rangle_s\}$ onto the states $\{|\gamma_i\rangle_s\}$ is $P_\gamma=1-P_\beta$. We have denoted the inner product $\langle \gamma_1|\gamma_2\rangle$ by $\gamma$. In this way, the mapping enables us to transform a pair of states with a given inner product $\alpha$ into a new pair of states with an inner product $\beta$ fixed in advance. We could have chosen $p_1=p_2$, but our aim is to study the most general situation where $p_1\neq p_2$.
It is worth to emphasize that if the process $\alpha\rightarrow \beta$ is successful, then the probability that the system $s$ is in $|\beta_i\rangle_s$ is
\begin{equation}
\eta_i^\prime=\frac{\eta_ip_i}{P_\beta}, \quad i=1,2.
\label{etaprime}
\end{equation}
Our study centers in the optimization of the success probability $P_\beta$, however, a particular application of this scheme could require to focus on the probabilities $\eta_i^\prime$ instead of $P_\beta$. For instance, the requirement of that $\eta_1^\prime=\eta_2^\prime$ imposes conditions that not necessarily optimize $P_\beta$.

The unitarity of $U_{sa}$ constrains the values of the complex parameters $\alpha$, $\beta$, and $\gamma$  through the equations
\begin{subequations}%
\begin{eqnarray}%
\alpha&=&\sqrt{p_1p_2}\beta+\sqrt{(1-p_1)(1-p_2)}\gamma,
\label{ca}\\
\alpha^*&=&\sqrt{p_1p_2}\beta^*+\sqrt{(1-p_1)(1-p_2)}\gamma^*.
\label{cb}
\end{eqnarray}%
\label{c}%
\end{subequations}%
We resort to a polar decomposition of each inner product, that is, $\alpha=|\alpha|e^{i\theta_\alpha}$,
$\beta=|\beta|e^{i\theta_\beta}$ and $\gamma=|\gamma|e^{i\theta_\gamma}$. Eqs. (\ref{c}) are linearly independent if and only if the following conditions are fulfilled
\begin{subequations}
\begin{eqnarray}
&|\alpha|\neq0,\hspace{0.09in}|\beta|\neq0,\hspace{0.09in}|\gamma|\neq0,&\label{const2}\\
&\theta_\beta-\theta_\alpha\neq k\pi,\hspace{0.09in}
\theta_\gamma-\theta_\alpha\neq m\pi,\hspace{0.09in}
\theta_\gamma-\theta_\beta\neq n\pi,&
\label{const3}
\end{eqnarray}%
\label{subeqc}%
\end{subequations}%
with $k,m,n\in\mathbb{Z}$. We note that the roles of $\beta$
and $\gamma$ are similar, this means that the desired inner product
could be codified indistinctly
in the outcome sets $\{|\beta_i\rangle_s\}$ or
$\{|\gamma_i\rangle_s\}$.

A general analysis of the Eqs. (\ref{c}) leads us to distinguish
three cases, which we shall study separately. In the first
case we assume that all constrains (\ref{subeqc}) are satisfied.
In the second one, we lift one of the constrains (\ref{const2})
by assuming $\alpha=0$, which implies that
$\theta_\gamma-\theta_\beta=\pm\pi$, so that one of the conditions (\ref{const3}) do not hold.
In the third case
we lift all constrains (\ref{const3}). This is equivalent to
consider $\alpha$, $\beta$, and $\gamma$ all real and in the interval $[-1,1]$.
We stress that the case $\beta=0$ corresponds to the well studied
unambiguous state discrimination protocol \cite{2.2}.

\subsection{Complex and non-zero inner products}
\label{subsection21}

From Eqs. (\ref{c}) we get the following relations
\begin{equation}
p_1p_2=x^2, \quad{\rm with}\;
x=\frac{|\alpha|\sin(\theta_\gamma-\theta_\alpha)}{|\beta|\sin(\theta_\gamma-\theta_\beta)}
\label{X}
\end{equation}
and
\begin{equation}
(1-p_1)(1-p_2)=y^2, \quad{\rm with} \;
y=\frac{|\alpha|\sin(\theta_\beta-\theta_\alpha)}{|\gamma|\sin(\theta_\beta-\theta_\gamma)}.
\label{Y}
\end{equation}
Combining these two equations we obtain a linear equation relating both probabilities $p_1$ and $p_2$, that is
\begin{equation}
p_1+p_2=1+x^2-y^2.
\label{LinearP1P2}
\end{equation}
Replacing Eq. (\ref{LinearP1P2}) in Eq. (\ref{X}) we obtain a second degree polynomial for $p_2$
\begin{equation}
p_2^2-(1+x^2-y^2)p_2+x=0,
\label{POLYNOMIAL}
\end{equation}
whose solutions are
\begin{equation}
p_{\pm}\!=\frac{1\! +x^2\!-y^2}{2}\pm\!\frac{\left[1\!-2(x^2\!+y^2)\!+(x^2\!-y^2)^2\right]^{\frac{1}{2}}}{2}.
\label{p1}
\end{equation}
It can be shown that solutions $p_{\pm}$ of Eq. (\ref{p1}) are real and constrained to the interval $[0,1]$ when the inner products $\alpha$, $\beta$ and $\gamma$ are such that
\begin{equation}
|x|+|y|\leq1.
\label{const1}
\end{equation}
Inserting $p_2=p_{\pm}$ into Eq. (\ref{POLYNOMIAL}) we obtain that the possible solutions to Eqs. (\ref{c}) are
\begin{equation}
p_1=p_+~{\rm and}~p_2=p_- \quad{\rm or}\quad p_1=p_-~{\rm and}~p_2=p_+.
\label{pmaspmenos}
\end{equation}
Assuming without lost of generality that
$|\gamma|\geq|\beta|$,  the constrain given by Eq. (\ref{const1}) indicates that the absolute values of the inner products must satisfy one of the two following inequalities
\begin{equation}
|\alpha|<|\beta|\leq|\gamma|\hspace{0.2in}\text{or}\hspace{0.2in}|\beta|<|\alpha|\leq|\gamma|,
\label{abg}
\end{equation}
in order to obtain physically acceptable solutions for $p_1$ and $p_2$. According to these inequalities it is
not possible to get inner products smaller than the initial
$|\alpha|$ in both outcomes. This conclusion is in agreement with the impossibility of the $1\rightarrow m$ deterministic cloning process, since
in this case we would have $\gamma=\beta=\alpha^{m+1}$ for $m\geq1$, which implies $|\gamma|=|\beta|=|\alpha|^{m+1}<|\alpha|$ for $m\geq1$, being in contradiction with Eqs. (\ref{abg}).
On the other hand, we have to stress that according to the conditions (\ref{abg}) it is permitted that $|\beta|=|\gamma|<|\alpha|$, this is, the absolute value of an inner product can be deterministically increased.
This is a remarkable result when the phase of the outcome inner product it is not relevant. We notice that
the phases $\theta_\beta$ and $\theta_\gamma$ can not be equal, as follows from the conditions (\ref{const3}).
Moreover, they are restricted by the constrain (\ref{const1}). It is worth to emphasize that it is allowed the process $\alpha\rightarrow |\beta|=|\gamma|=1$, what means that the initial information codified in the different states $|\alpha_i\rangle$ can be deterministically deleted.

Replacing both solutions (\ref{pmaspmenos}) with (\ref{p1}) in Eq. (\ref{pbeta})
we obtain two solutions $P_{\beta,\pm}$ for the success probability of mapping the states $\{|\alpha_i\rangle\}$ onto the states $\{|\beta_i\rangle\}$, namely
\begin{equation}
P_{\beta,\pm}\!=\frac{1\!+\!x^2\!-\!y^2}{2}\pm\frac{|\eta_1\!-\!\eta_2|\!\left[1\!-\!2(\!x^2\!+\!y^2\!)\!+\!(\!x^2\!-\!y^2\!)^2\right]}{2}^{\frac{1}{2}}\!.
\label{pbeta2}
\end{equation}
We observe that $P_{\beta,+}= P_{\beta,-}$  when the a priory probabilities $\eta_1$ and $\eta_2$ are equal.
These probabilities are also equal when the equality in Eq. (\ref{const1}) is satisfied, independently of the values of $\eta_1$ and $\eta_2$. In general, the maximum probability is $P_{\beta,+}$ and the minimum
one is $P_{\beta,-}$. When $P_\beta$ is maximum,
$P_\gamma=1-P_\beta$ will be minimum, and vice versa.

It is important to note that both $P_\beta$ and $P_\gamma$ depend on
the phase differences $|\theta_\beta-\theta_\alpha|$,
$|\theta_\gamma-\theta_\alpha|$, and $|\theta_\gamma-\theta_\beta|$. Thus,
these quantities can be used to increase the probability of a particular mapping. This effect can be visualized in the particular case in
which $\theta_\beta=\theta_\alpha+\delta$, with $|\delta|\ll1$.
Assuming a Maclaurin serie of $P_{\beta,+}$ up to first order in
$\delta$ we obtain
\begin{eqnarray}
P_{|\beta|e^{i(\theta_\alpha+\delta)},+}&=&\frac{1}{2}\left(1+\frac{|\alpha|^{2}}{|\beta|^{2}}\right)
+\frac{|\eta_{1}-\eta_{2}|}{2}\left\vert1-\frac{|\alpha|^{2}}{|\beta|^{2}}\right\vert  \nonumber\\
&&-\frac{|\alpha|^{2}}{|\beta|^{2}}\frac{1\pm|\eta_{1}-\eta_{2}|}
{\tan(\theta_{\alpha}-\theta_{\gamma})}\delta+O\left(\delta^{2}\right),\nonumber
\end{eqnarray}
where the plus (minus) sign has to be considered when
$|\alpha|>|\beta|$ ($|\alpha|<|\beta|$). From this expression it is
clear that the success probability could be increased or decreased
depending on the sign of $\delta=\theta_\beta-\theta_\alpha$ and on $\theta_\gamma$ through the
function $\tan(\theta_\alpha-\theta_\gamma)$.

\begin{figure}[t]
\includegraphics[width=0.34\textwidth,angle=-90]{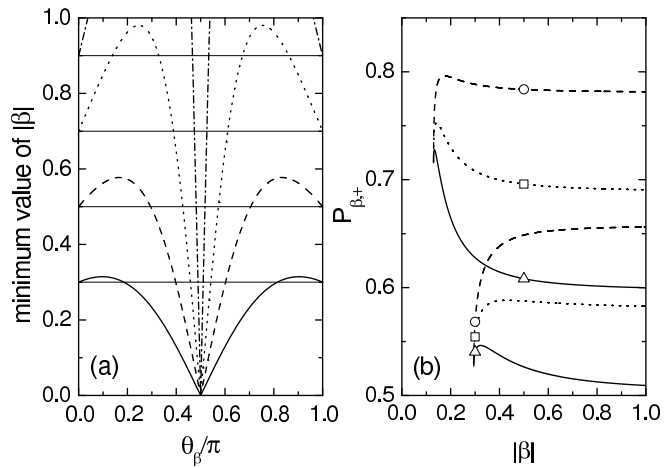}
\caption{(a) The minimum allowed value of $|\beta|$ as a function
of $(\theta_\alpha-\theta_\beta)/\pi$ for $\alpha=0.3$ (solid),
$\alpha=0.5$ (dashes), $\alpha=0.7$ (dots), and $\alpha=0.9$ (dash
dot). The horizontal grey lines correspond to the respective
$|\alpha|$ value. (b) The maximal success probability
$P_{\beta,+}$ as a function of $|\beta|$ for $\alpha=0.3$ (black) and $\alpha=0.5$
(grey). We consider $|\eta_1-\eta_2|=0.3$ (solid),
$|\eta_1-\eta_2|=0.7$ (dashes), and $|\eta_1-\eta_2|=0.5$ (dots). In both figures we have considered $\gamma=\pm
ie^{i\theta_\beta}$, $\theta_\beta=0.6\pi$ and $\theta_\alpha=0$.}
\label{fig1}
\end{figure}

Returning to the general case where we use this mapping to change
the initial inner product $\alpha$ to a desired $\beta$, $\gamma$ can be used to increase the optimal
probability $P_{\beta,+}$. This probability reaches
its highest value when $\gamma=\pm ie^{i\theta_\beta}$, independently of the value of
$|\eta_1-\eta_2|$, which means that the process
$\alpha\rightarrow\beta$ has a larger probability when $|\gamma_1\rangle=\pm
ie^{i\theta_\beta}|\gamma_2\rangle$, that is, when both states are in the same ray. The cost of maximizing the probability of success this way
is that, if the process fails, it is not possible to reverse the mapping, not even probabilistically, and the information about the initial states $\{|\alpha_i\rangle_s\}$ is totally lost. In this case Eq. (\ref{const1}) indicates that the inner product $\beta$ is lower bounded as
\begin{equation}
|\beta|\geq\frac{|\alpha||\cos(\theta_\alpha-\theta_\beta)|}{1-|\alpha||\sin(\theta_\alpha-\theta_\beta)|}.
\label{bb}
\end{equation}
Fig. \ref{fig1}.a shows the lower bound of $|\beta|$ given by Eq. (\ref{bb}) as a function of $\theta_\beta/\pi$ for different values of $\alpha$ ($\theta_\alpha=0$). We see that, depending on $\alpha$ and of the phase $\theta_\beta$, the value of $|\beta|$ can be lower, equal, or higher than $\alpha$ or simply there is no solution. We point out that, as follows from the constrain (\ref{const3}) and since $\theta_\beta-\theta_\gamma=\pm\pi/2$, the values $\theta_\alpha-\theta_\beta=\pm\pi/2$ are forbidden.
Fig. \ref{fig1}.b shows the maximal success probability
$P_{\beta,+}$ as a function of $|\beta|$ for $\alpha=0.3$ (black line) and $\alpha=0.5$
(grey line). We consider three different values of $|\eta_1-\eta_2|$ for each value of $\alpha$, namely $|\eta_1-\eta_2|=0.3$ (solid), $|\eta_1-\eta_2|=0.7$ (dashes), and $|\eta_1-\eta_2|=0.5$ (dots). From this figure we observe that the
probability of mapping from $\alpha=0.3$ to $|\beta|=0.5$ (black:
circle, square, and triangle) is higher than the
probability of mapping from $\alpha=0.5$ to $|\beta|=0.3$ (grey:
circle, square, and triangle). Clearly, the optimal probability depends on the direction of the mapping. We also notice
that the optimal probability depends on the phase
$\theta_\beta$, which permits improving the success probability of the mapping.

On the other hand, it can be shown straightforwardly from Eq. (\ref{const1}) that it is possible to increase the range of values of $|\beta|$ by setting $|\gamma|<1$ at the expense of decreasing the maximum success probability of the mapping.

\subsection{Non-orthogonal pure states preparation}
\label{subsection22}

The second case arises under the condition that $\alpha$ vanishes. This particular mapping generates probabilistically non-orthogonal states with inner product $\beta$ or $\gamma$ from two orthogonal states. It follows from Eq. (\ref{ca}) that
$\theta_\gamma-\theta_\beta=\pm\pi$ and consequently the constrain Eq. (\ref{cb}) is also lifted. Thereby, both Eqs. (\ref{c}) are reduced to the single equation
\begin{equation}
0=\sqrt{p_1p_2}|\beta|-\sqrt{(1-p_1)(1-p_2)}|\gamma|.
\label{NEWEQUATION}
\end{equation}
Assuming that $|\beta|$ and $|\gamma|$ are both different from zero,  Eq. (\ref{NEWEQUATION}) enables us to obtain $p_2$ as a function of $p_1$,
\begin{equation}
p_2=\frac{(1-p_1)|\gamma|^2}{p_1|\beta|^2+(1-p_1)|\gamma|^2}.
\label{p2}
\end{equation}
We see that $p_2$ is always in the interval $[0,1]$. Hence, the mapping of states $\{|\alpha_i\rangle\}$ initially orthogonal onto states $\{|\beta_i\rangle\}$ with inner product $\beta$ has the total success probability $P_{\beta,\alpha=0}$ given by
\begin{equation}
P_{\beta,\alpha=0}=\eta_1p_1+\eta_2\frac{(1-p_1)|\gamma|^2}{p_1|\beta|^2+(1-p_1)|\gamma|^2},\label{pbeta2b}
\end{equation}
which is a function of $p_1\in [0,1]$.
This probability can be optimized with respect to this parameter. The derivative $dP_{\beta,\alpha=0}/dp_1$ vanishes when $p_1$ takes the value
\begin{equation}
\bar{p}_1=\frac{1-\frac{|\beta|}{|\gamma|}\sqrt{\frac{\eta_2}{\eta_1}}}{1-\frac{|\beta|^2}{|\gamma|^2}}.
\label{P1OPTIMAL}
\end{equation}
Replacing Eq. (\ref{P1OPTIMAL}) in Eq. (\ref{pbeta2b}) we get
the extreme success probability
\begin{equation}
\bar{P}_{\beta,\alpha=0}=\frac{1-2\frac{|\beta|}{|\gamma|}\sqrt{\eta_1\eta_2}}{1-\frac{|\beta|^2}{|\gamma|^2}}.
\label{PBETAALPHA0}
\end{equation}
Considering the constrain $\bar{p}_1\in [0,1]$, Eq. (\ref{PBETAALPHA0}) corresponds to a maximum when
\begin{equation}
|\beta|<|\gamma|\quad \mbox{and} \quad \frac{|\beta|}{|\gamma|}<\min\left\{\sqrt{\frac{\eta_1}{\eta_2}},\sqrt{\frac{\eta_2}{\eta_1}}\right\},
\label{interval1}
\end{equation}
and to a minimum when
\begin{equation}
|\beta|>|\gamma|\quad \mbox{and} \quad \frac{|\beta|}{|\gamma|}>\max\left\{\sqrt{\frac{\eta_1}{\eta_2}},\sqrt{\frac{\eta_2}{\eta_1}}\right\}.
\label{interval2}
\end{equation}
 We note that the probability $\bar{P}_{\beta,\alpha=0}$ is monotonously decreasing as a function of $|\beta|/|\gamma|$. Then,
 when the condition (\ref{interval1}) is fulfilled, we can increase the maximum probability $\bar{P}_{\beta,\alpha=0}$ by making $|\gamma|=1$, this is, the information codified in the states $\{|\alpha_i\rangle\}$ is lost if the process $\alpha=0\rightarrow \beta$ fails.
 When (\ref{interval2}) is satisfied,  the function (\ref{pbeta2b}) is concave as a function of $p_1$, then its maximum value is in one of the borders of the interval $p_1 \in [0,1]$. In this case $P_{\beta,\alpha=0}=\eta_2$ for $p_1=0$ and $P_{\beta,\alpha=0}=\eta_1$ for $p_1=1$,
so that the maximum value of $P_{\beta,\alpha=0}$ depends on the highest a priory probability.
We have to note that, although the success probability is optimum,  when $p_1=0$ the probability $\eta_1^\prime$ vanishes, which means that the system is never mapped to $|\beta_1\rangle$.

If $|\beta|$ and $|\gamma|$ do not fulfill any of Eqs. (\ref{interval1}) and (\ref{interval2}), the total success probability
$P_{\beta,\alpha=0}$ of Eq. (\ref{pbeta2b}) is monotonously increasing with $p_1$ if $\eta_1>\eta_2$ or monotonously decreasing if $\eta_1<\eta_2$, so the maximum value of $P_{\beta,\alpha=0}$ is $\eta_1$ or $\eta_2$, respectively.

\begin{figure}[h]
\includegraphics[width=0.30\textwidth,angle=-90]{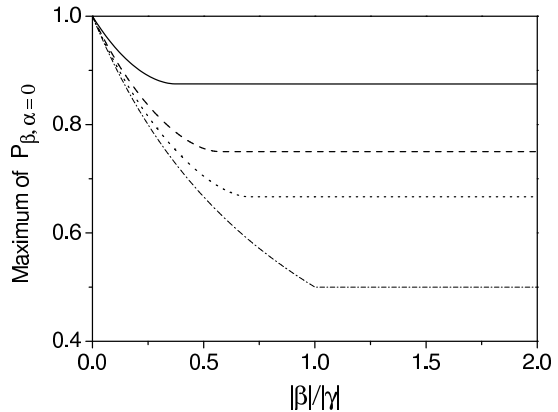}
\caption{Maximum success probability $P_{\beta,\alpha=0}$ of preparing non-orthogonal states as a function of $|\beta|/|\gamma|$ for different values of $\eta_1$: $\eta_1=1/8$ (solid), $\eta_1=1/4$ (dashes), $\eta_1=1/3$ (dots),  and $\eta_1=1/2$ (dash dot).}
\label{pa0}
\end{figure}

The above analysis is summarized in Fig. \ref{pa0}, which shows the maximum success probability as a function of $|\beta|/|\gamma|$ for different values of $\eta_1$.
This function is monotonously decreasing with $|\beta|/|\gamma|$ between $0$ and $\sqrt{\eta_</\eta_>}$, and it remains constant and
equal to $\eta_>$ for larger values of $|\beta|/|\gamma|$. We have put $\eta_<=\min\{\eta_1,\eta_2\}$ and
$\eta_>=\max\{\eta_1,\eta_2\}$.

Let us consider the particular case where $|\beta|=|\gamma|$. Here we map
two orthogonal states onto other two states with inner product either
$\beta$ or $-\beta$. Thus, we deterministically prepare two states with
arbitrary inner product module $|\beta|$ starting from two
orthogonal states.
This result is even valid for $|\beta|=|\gamma|=1$ which means that the
initial information codified in the orthogonal states
$\{|\alpha_i\rangle\}$ can be deleted with unitary probability $1$.

\subsection{Conditioned phases}
\label{subsection23}

Let us analyze the third case, where we lift the constrain Eq.
(\ref{const3}). We assume that $\theta_\beta-\theta_\alpha= k\pi$. This, together with Eqs. (\ref{c}) lead to
$\theta_\gamma-\theta_\alpha=m\pi$ and $\theta_\gamma-\theta_\beta=
(m-k)\pi$, with $k$, $m$ $\in\mathbb{Z}$. Under these conditions the
two equations (\ref{c}) are linearly dependent and they are reduced
to one, with $\alpha$, $\beta$, and $\gamma$ real and in the interval $[-1,1]$.

From Eq. (\ref{ca}) we obtain $p_2$ as a function of $p_1$
\begin{equation}
\small
p_{2,\pm}=\frac{\left(|\alpha||\beta|\sqrt{p_{1}}
\pm|\gamma|\sqrt{\left(1-p_{1}\right)\left[\gamma^{2}-\alpha^{2}-\left(\gamma^{2}
-\beta^{2}\right)p_{1}\right]}\right)^{2}}{\left[\gamma^{2}-\left(\gamma^{2}-\beta^{2}\right)p_{1}\right]^{2}}.
\label{p2real}
\end{equation}
Without loss of generality, we assume that
$|\beta|\leq|\gamma|$. In addition to the requirement $0\le p_{2,\pm}\leq1$,
in order to guarantee that these probabilities are real, the following
inequality must be satisfied
\begin{equation}
0<p_{1}\leq\min\{\frac{\gamma^2-\alpha^2}{\gamma^2-\beta^2},1\}. \label{ine1}
\end{equation}
This is fulfilled in the two following cases: i)
$|\gamma|\ge|\beta|>|\alpha|$ and ii) $|\gamma|>|\alpha|\ge|\beta|$, so that the outcomes modules can not be both less than $|\alpha|$.
\begin{figure}[t]
\includegraphics[width=0.30\textwidth,angle=-90]{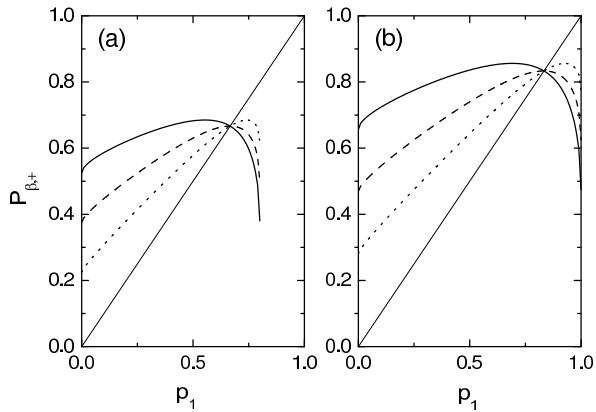}
\caption{The success probabilities $P_{\beta,+}$
as a function of $p_1$ for different values of $\eta_1$, for
(a) $\alpha=1/3$, $\beta=1/6$, and $\gamma=2/3$, (b) $\alpha=1/6$,
$\beta=1/3$, and $\gamma=2/3$. In both figures $\eta_1=0.3$ (solid), $\eta_1=0.5$ (dashes), and $\eta_1=0.7$ (dots).
The gray line correspond to $p_1=p_2$.}
\label{fig2}
\end{figure}

Using the solutions (\ref{p2real}) we get two success probabilities for the process $\alpha\rightarrow\beta$,
\begin{equation}
P_{\beta,\pm}=\eta_1p_1+\eta_2p_{2,\pm}.
\end{equation}
It is evident that the optimum one is $P_{\beta,+}$. Fig. \ref{fig2} shows $P_{\beta,+}$ as a function of $p_1$
for fixed $\alpha,\beta$ and $\gamma$ and different values of $\eta_1$. In Fig. \ref{fig2}.a the desired outcome $\beta$ is less than
$\alpha$, whereas in Fig. \ref{fig2}.b $\beta$ is larger than $\alpha$. In both pictures we observe that the value of $p_1$ for which
$P_{\beta,+}$ is maximum depends on $\eta_1$ and only when $\eta_1=1/2$ (equal a priory probabilities) this occurs when $p_1=p_2$. This
justifies the choice of $p_1$ in general different of $p_2$ in the process described by Eqs. (\ref{TRANSFORMATION}).
On the other hand, we also see that the optimal success probability for decreasing the inner product is smaller than the one for
increasing it. That is explained by the fact that in the process described by Fig. \ref{fig2}.a information is gained if the process is successful, while in the process of Fig. \ref{fig2}.b, information is partially lost.

Finally, according to the condition (i) $|\beta|=|\gamma|>|\alpha|$ is allowed. In this case $\theta_\beta$ can be equal to $\theta_\gamma$, so that the inner product can be deterministically modified, but with the restriction
$\theta_\beta=\theta_\alpha+k\pi$. The special case with $\beta=\gamma=1$ gives account of deleting the information stored in two quantum states.

\section{Comparison with unambiguous state discrimination}
\label{sectionA}

In two of the previous subsections we analyzed the mapping between sets of non-orthogonal states.
A different scheme for implementing this mapping would be the concatenation of unambiguous state discrimination
of the initial states $\{|\alpha_i\rangle_s\}$ with a unitary preparation of the states $\{|\beta_i\rangle_s\}$.
Clearly, this scheme allows us to know with certainty the initial state which will be mapped and the value of $\beta$ is not constrained. The success probability of this alternative scheme corresponds to the success probability $P_{USD}$ of unambiguous state discrimination, which given by \cite{2.2}
\begin{equation}
P_{USD}=1-2\sqrt{\eta_1\eta_2}|\alpha|.\label{pp}
\end{equation}
Next we compare this with the probability given by Eq. (\ref{pbeta2}), and we show the importance of the complex nature of the involved inner products $\alpha$, $\beta$ and $\gamma$ for get larger success probabilities. In what follows we put $\theta\equiv\theta_\beta-\theta_\alpha$.

\begin{figure}[t]
\includegraphics[width=0.37\textwidth,angle=-90]{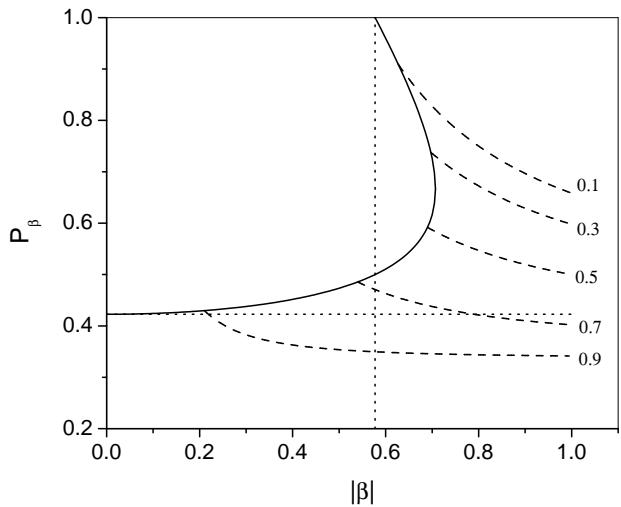}
\caption{$P_{|\beta|_{min}}$ as a function of $|\beta|_{min}$ for
$\gamma=\pm ie^{\theta_\beta}$ and $|\alpha|=1/\sqrt{3}$ (solid).
Success probability $P_{\beta,+}$ as a function of $|\beta|$ for the
same $|\alpha|$, $\eta_1=\eta_2$, and different values of
$\theta=f\pi/2$ with the values of $f$ given at the right hand of
each dashes curves. The horizontal dotted line corresponds to
$P_{USD}=1-|\alpha|$, and the vertical dotted line corresponds to
$|\beta|=|\alpha|$.} \label{Comp}
\end{figure}

When the constrains (\ref{subeqc}) are all satisfied, according to the inequality (\ref{bb}), when
$\gamma=\pm ie^{i\theta_\beta}$ and
for a given inner product $\alpha$
there is a minimum value of $|\beta|$ given by
\begin{equation}
|\beta|_{min}=\frac{|\alpha||\cos\theta|}{1-|\alpha||\sin\theta|}.
\label{const4}
\end{equation}
Fig. \ref{Comp} shows $P_\beta$ as a function of $|\beta|$, for $\eta_1=\eta_2$ and different values of $\theta$.
The numbers labeling the curves corresponds to factors multiplying $\pi/2$.
The solid curve correspond to the value of $P_\beta$ evaluated in $|\beta|_{min}$, which is given by
\begin{equation}
P_{|\beta|_{min}}=1-\frac{|\beta|_{min}^2\pm\sqrt{|\alpha|^2(1+|\beta|_{min}^2)-|\beta|_{min}^2}}{1+|\beta|_{min}^2}.
\label{eq-7}
\end{equation}
The minus sign correspond to  $0<\theta\leq \arccos{(\alpha)}$ and the plus sign to   $\arccos{(\alpha)}\leq \theta< \pi/2$. These features reproduces for $\pi/2<\theta< \pi$.
In the figure, these probabilities correspond to the solid curve. The dash lines correspond to the probability $P_{\beta,+}$ of Eq. (\ref{pbeta2}) as a function of $|\beta|$ for different values of $\theta$, for $|\alpha|=1/\sqrt{3}$ (vertical dotted line). We have considered $\theta=f\pi/2$, where $f$ is indicated for each of the curves in the figure. The horizontal dotted line is the corresponding probability $P_{USD}$, Eq. (\ref{pp}).

The figure shows clearly that for any target $|\beta|$ it is always possible to choose a phase difference $\theta$ where the probability $P_{\beta}$ is higher than $P_{USD}$. This is particularly interesting when the scheme is used to obtain a target $|\beta|<|\alpha|$ and the phase does not matter. If the aim is increasing the absolute value of the inner product, we have shown that this can be done deterministically by choosing $|\gamma|=|\beta|$.
We have to note that due to the constrain (\ref{const2}) the proposed scheme does not allows us to reach
exactly the value $|\beta|=0$, but we can approach to it with a probability tending to
$P_{USD}=1-|\alpha|$.

It is direct to show analytically that  the probability (\ref{eq-7})
is always higher than $P_{USD}$ when $\eta_1=\eta_2$, for any
$\alpha$. This means that there exists a family of targets $\beta$
with probabilities higher than the respective $P_{USD}$. For any
other target $\beta$, which of the schemes gives a higher success
probability has to be examined in that particular case.

\section{Discussion}
\label{application}

An application of this scheme emerges when in the Eqs.
(\ref{TRANSFORMATION}) the
states $\{|\alpha_i\rangle_s\}$ are replaced by the states
$\{|\alpha_i\rangle_s\otimes|\omega\rangle^{\otimes m}\}$, where $|\omega\rangle^{\otimes m}$
is a state of a $m-$partite quantum system, and the states
$\{|\beta_i\rangle_s\}$ are replaced by
$\{|\alpha_i\rangle^{\otimes(m+1)}\}$. This process describes
a probabilistic $1\rightarrow m+1$ cloning machine of the two linearly
independent states $\{|\alpha_i\rangle_s\}$. Thereby, we get the
optimal probability of generating successfully $m+1$ copies of any pair of states
$\{|\alpha_i\rangle_s\}$ by replacing $\beta=\alpha^{(m+1)}$ in
Eqs. (\ref{c}). As discussed in subsections \ref{subsection21} and \ref{subsection23}, the
success probability reaches its highest value when
the information is totally lost if the process fails, that is, when
$\gamma=\pm ie^{i(m+1)\theta_\alpha}$ and $|\gamma|=1$, respectively.

Finally, as we mentioned before, a deterministic quantum deleting scheme can be achieved
by choosing the two possible outcomes consisting of parallel states.

\section{Summary}
\label{summary}

In summary, we have proposed and analyzed a scheme which maps conclusively a couple of known pure states onto
another couple, allowing changing probabilistically the inner product on demand. The scheme is performed by
a bipartite unitary transformation on the system and an ancillary system, followed by a von Neumann measurement.
This assumes different a priory probabilities, as well as different projection probabilities
into the target states via the ancillary states. The latter probabilities are introduced by the unitary transformation and
the fact that they are different permits in general optimize the success probability of the mapping.
Our analysis shows that the phases of the involved inner products play an important role in the
increase of the success probability of the desired process.

We have compared this scheme with a process of mapping via unambiguous
states discrimination, obtaining larger success probabilities.
In the analyzed case, our scheme does not allow us to reach
exactly the orthogonality, but we can approach to it with a probability
which is always larger than $P_{USD}$.

\begin{acknowledgments}
This work was supported by Grants: Milenio ICM P06-067F and FONDECyT
N$^{\text{\underline{o}}}$ 1080535, 1080383 and 1080660.

\end{acknowledgments}

\end{document}